\documentclass[a4paper,11pt]{article}
\usepackage{pos}

\def\jraa{R_{\tiny{\rm AA}}}
\def\aSC{{\kappa_{\rm sc}}}

\title{Jet suppression from small to large radius}
\author*[a]{Daniel Pablos}

\affiliation[a]{Department of Physics and Technology, University of Bergen, 5007 Bergen, Norway}

\emailAdd{daniel.pablos@uib.no}

\abstract{The angular dependence of jet suppression encodes key information about the process of energy and momentum hydrodynamization, and for this reason can be used to greatly improve our understanding of fundamental aspects of the jet/QGP interaction. In this work we study jet suppression from small to very large radius, for low and very high energy jets at the LHC and RHIC. We use the hybrid strong/weak coupling model for jet quenching that combines perturbative shower evolution with an effective strongly coupled description of the energy and momentum transfer from the jet into the QGP. Because of energy-momentum conservation, the wake created by the jet enhances or depletes the yield of particles generated at the freeze-out hypersurface depending on their orientation with respect to the direction of the jet. We find that jet suppression is remarkably independent of the anti-$k_T$ radius R, specially at LHC, first slightly increasing by opening R, then at larger values of R slowly decreasing. This nearly independence of jet suppression with R arises from two competing effects, namely the larger energy loss of the energetic jet components, which tends to increase suppression, against the partial recovery of the lost energy due to medium response, which reduces suppression. We find that the boosted medium from the recoiling jet depletes the amount of QGP in the direction opposite to it in the transverse plane, inducing energy loss due to an over-subtraction effect. We show that this unique signature of the hydrodynamization of part of the jet energy can be scrutinized by selecting samples of dijet configurations with different relative rapidities between the leading and the subleading jet.}

\FullConference{%
  HardProbes2020\\
  1-6 June 2020\\
  Austin, Texas}

\begin{document}
\maketitle

\section{Introduction}
Jets are collimated sprays of hadrons that originate from the evolution of highly virtual partons created in a hard scattering. These hard probes are modified through their passage through the medium created in heavy-ion collisions, a phenomenon known as jet quenching~\cite{Mehtar-Tani:2013pia,Qin:2015srf}. The study of such modifications gives unique access to the inner workings of this medium, the quark-gluon plasma (QGP), an exploding fireball that is very well described by relativistic hydrodynamic simulations~\cite{Acharya:2018zuq}.

Colored energetic partons can interact with the QGP and will ``lose'' energy through a set of mechanisms, e.g. stimulated radiation or elastic collisions. It is clear that some of the ``lost'' energy will reach the medium scale, characterised by its temperature $T$, contributing into having a little more, and a little hotter QGP in a given region. How much, and how fast a certain amount of energy hydrodynamizes will depend on details of the QGP such as the strength of the coupling between the jet and the medium and the nature of the degrees of freedom of the plasma. Given the phenomenologically supported evidence of the fluid nature of the QGP, it is logical to expect that the thermalised ``lost'' energy, and momentum, will also display a characteristic fluid like behaviour. In this work we will study the implications of the hydrodynamization of energy and momentum on the angular dependence of jet suppression. We will also see that the modification of the QGP background will induce long-range correlations between the wakes excited by each of the jets of a dijet system, a novel observation that can be quantified by studying dijet configurations with different rapidity gaps.

\section{The hybrid strong/weak coupling model}
High energy partons start with an evolution scale $Q\sim p_T$ which is relaxed through successive splittings as described by the DGLAP evolution. This stochastic evolution governs the dynamics of parton showers, which originate the jets. While the parton shower is being formed, it is also propagating within a medium with a temperature scale $T\sim \Lambda_{\textrm{QCD}}$. This complex system has the advantage of presenting two widely separated scales, namely $Q \gg T$, motivating the convenience of adopting a hybrid description: even though parton splittings can be well described by perturbative QCD, the interaction of the jet constituents with the plasma is assumed to be dominated by non-perturbative processes. This is the basis of the hybrid strong/weak coupling model \cite{Casalderrey-Solana:2014bpa,Casalderrey-Solana:2015vaa}, in which the perturbative splittings are simulated with PYTHIA8~\cite{Sjostrand:2014zea} and the non-perturbative interaction is modelled through insights from the gauge/gravity correspondence. The energy loss rate in a strongly coupled $\mathcal{N}=4$ SYM plasma at large $N_c$ reads \cite{Chesler:2014jva,Chesler:2015nqz}

\begin{equation}\label{CR_rate}
\left. \frac{d E_{}}{dx}\right|_{\rm strongly~coupled}= - \frac{4}{\pi} E_{\rm in} \frac{x^2}{x_{\rm therm}^2} \frac{1}{\sqrt{x_{\rm therm}^2-x^2}} \quad , 
\end{equation}

where $E_{\rm in}$ is the initial energy of the parton and $x_{\rm \small therm}=(E_{\rm in}^{1/3}/T^{4/3})/2 \aSC$ is the thermalization distance, were $T$ is the local temperature of the plasma as read from event-averaged hydrodynamic profiles \cite{Shen:2014vra}. The quantity $\aSC$, which depends on the 't Hooft coupling differently depending on how the holographic system is prepared, is taken as a free parameter that we fit to jet and hadron data \cite{Casalderrey-Solana:2018wrw}. The deposited energy and momentum into the plasma will originate a wake~\cite{Chesler:2007an}, which by energy-momentum conservation has to be correlated with the direction of the jet. We estimate the impact on the final hadronic spectrum that the decay of such wake induces at the freeze-out hypersurface by solving the Cooper-Frye spectrum~\cite{Cooper:1974mv} to first order, assuming small perturbations on top of a Bjorken fluid which stay localised around the rapidity of the jet. The result is~\cite{Casalderrey-Solana:2016jvj}

\begin{equation}
\label{onebody}
\begin{split}
E & \frac{d\Delta N}{d^3p}=\frac{1}{32 \pi} \, \frac{m_T}{T^5} \, \cosh(y-y_j)  \exp\left[-\frac{m_T}{T}\cosh(y-y_j)\right] \\
 &\times \Bigg\{ p_{T} \Delta P_{T} \cos (\phi-\phi_j) +\frac{1}{3}m_T \, \Delta M_T \, \cosh(y-y_j) \Bigg\} \quad ,
\end{split}
\end{equation}
where $m_T$, $\phi$ and $y$ are the transverse mass, azimuthal angle and rapidity of a given hadron, respectively, and where $\phi_j$, $y_j$, $\Delta P_{T}$ and $\Delta M_T$ are the azimuthal angle, rapidity, deposited transverse momentum and deposited transverse mass from the jet, respectively. A noteworthy feature of this expression is that it becomes negative for large azimuthal separations. The physical interpretation is that a boosted fluid cell will tend to emit less particles in the direction opposite to the boost, which is the direction of the jet. This is a manifestation of the diffusion wake, which yields a reduction of the amount of QGP in a given region of phase space with respect to an unperturbed background. This depletion of the background has also been observed for full non-linear hydrodynamical simulations~\cite{Chen:2017zte} and semi-analytical approaches~\cite{Yan:2017rku}. In analogy to the shape of a wave on the surface of a liquid, we can call the positive, near-side contribution from Eq.~\eqref{onebody} the QGP ``ridge'', or crest, and the negative, away side contribution the QGP trough. Non-hydrodynamized partons fragment into hadrons through the Lund model included in PYTHIA8.

\section{Jet suppression and the effect of the wake}

\begin{figure*}[t!]
\includegraphics[width=1\textwidth%,height=0.62\textheight
]{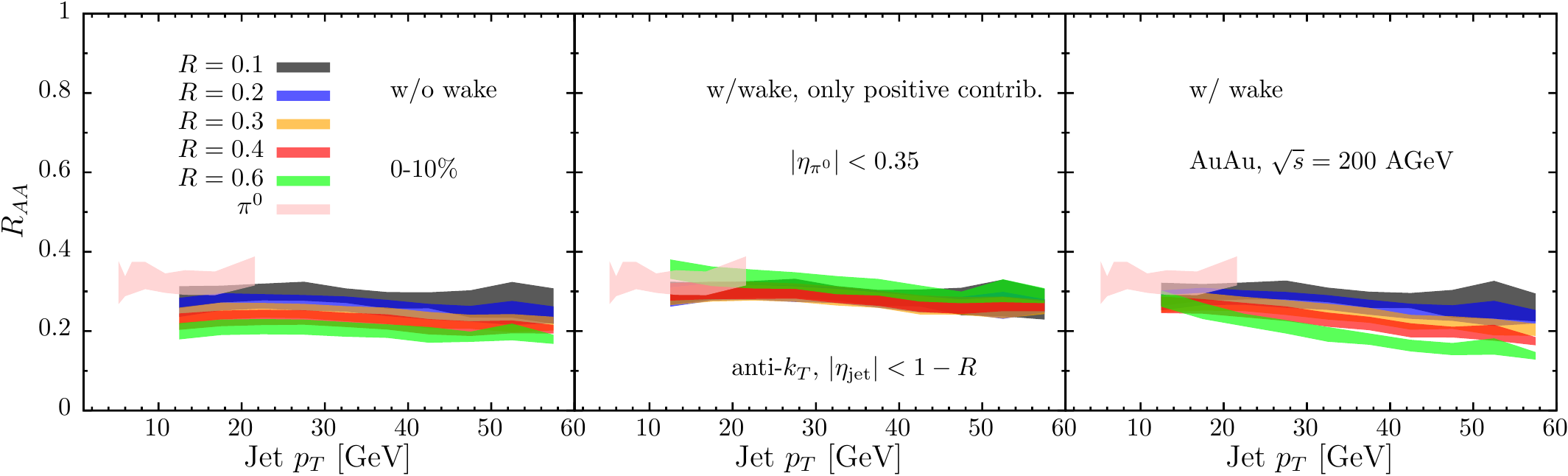}
\vspace{-0.2in}
\caption{\label{fig:raavsr} The anti-$k_T$ jet radius $R$ dependence of $\jraa$ in AuAu collisions at $\sqrt{s}=200$ AGeV for the $0-10$\% centrality class. The left (middle) panel only contains the non-equilibrium (non-equilibrium + QGP ridge) contribution, while the full result that includes the effects of the QGP trough is shown in the right panel.}
\end{figure*}

We show in Fig.~\ref{fig:raavsr} the results for jet suppression at RHIC, for AuAu collisions at $\sqrt{s}=200$ AGeV in the 0-10\% centrality class (for the results at LHC see~\cite{Pablos:2019ngg}). Jets are reconstructed using FastJet~\cite{Cacciari:2011ma} with the anti-$k_T$ algorithm for different radius $R$. In the left panel we show results that do not include the contribution of the wake. Jet suppression increases with $R$ because wider jets contain more energy loss sources, and are more quenched~\cite{Milhano:2015mng,Rajagopal:2016uip,Casalderrey-Solana:2019ubu,Caucal:2020xad}. The middle panel only includes the positive contribution of the wake, the QGP ``ridge''. By opening the jet cone, more of the deposited energy can be recovered, competing against the previous effect and yielding a relatively flat dependence of jet suppression for different $R$. However, if we include all contributions, which corresponds to the only physical, energy-momentum conserving scenario in our model, we observe that larger $R$ jets are again more suppressed than narrower ones. Wider jets are more likely to capture the contribution from the QGP trough, which actually comes from the wake originated by the recoiling jet. At RHIC energies, for high jet $p_T$ there is barely any phase space left to have the dijet situated away from mid-rapidity, so that in particular the rapidities of the jets are very similar.  This is a necessary condition for the wake of one jet to affect the other, since the rapidity distribution of the particles from Eq.~\ref{onebody} is relatively narrow, $|\Delta y| \sim 1$. On the contrary, at LHC energies, even at high $p_T$ we can have dijet systems with a large rapidity gap, providing the right conditions to study the effect of the QGP trough more differentially.

\begin{figure}
\centering
\includegraphics[width=0.5\textwidth%,height=0.62\textheight
]{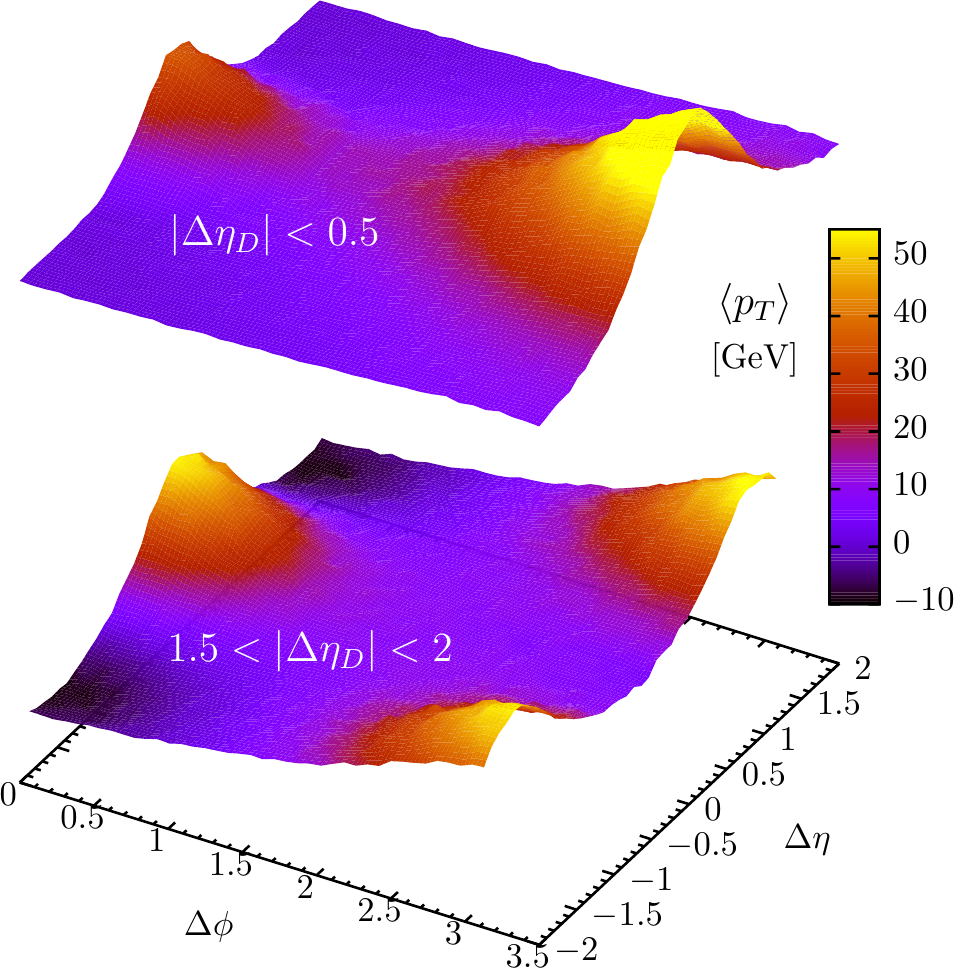}
\includegraphics[width=0.45\textwidth%,height=0.3\textheight
]{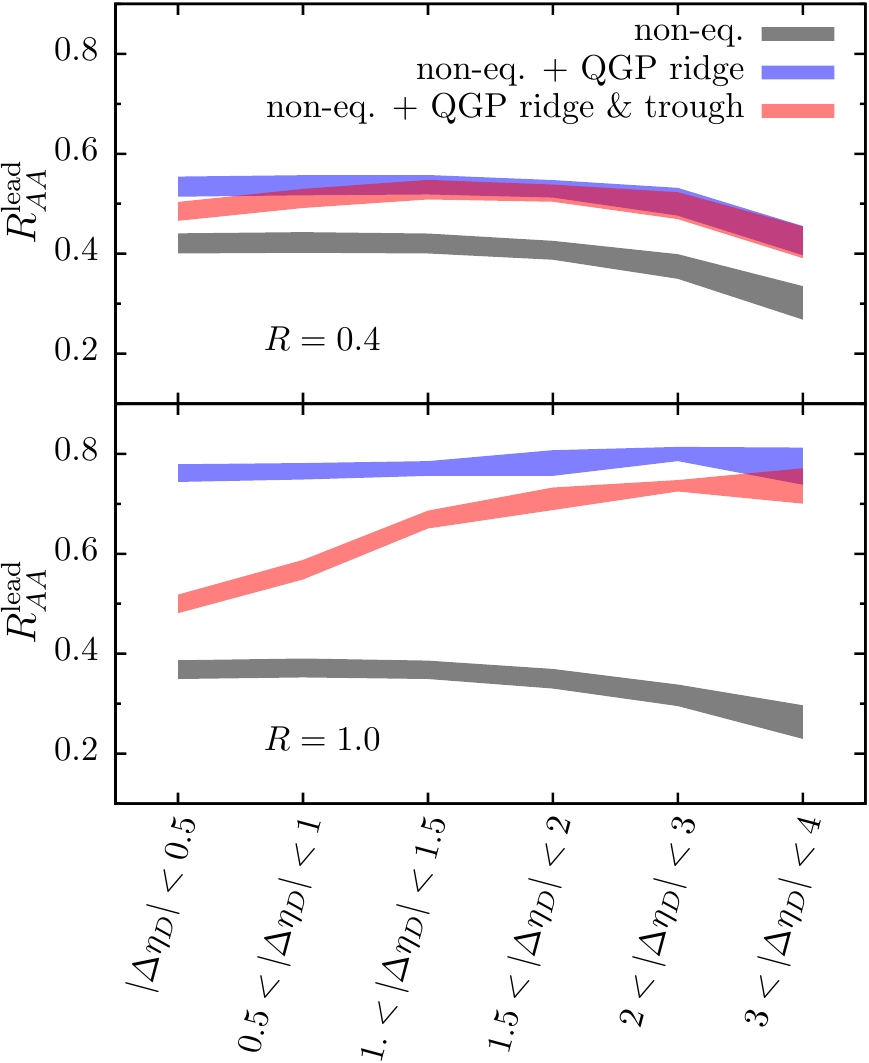}
\caption{\label{fig:deltaeta} \emph{Left}: the $\langle p_T \rangle$ density of the particles coming from the wake with respect to the leading jet direction, in terms of $\Delta \eta$ and $\Delta \phi$, for PbPb collisions at $\sqrt{s}=5.02$ ATeV. The leading jet has $p_T^L>250$ GeV and the subleading has $p_T^S>80$ GeV, with an angular separation in the transverse plane $\Delta \phi_D > 2\pi/3$ and an absolute difference in rapidity $|\Delta\eta_D|$. \emph{Right}: jet suppression $R_{AA}^{\rm lead}$ for leading jets above $p_T^L>250$ GeV, as a function of the pseudorapidity separation $|\Delta\eta_D|$ with respect to the back-to-back subleading jet with $p_T^S>80$ GeV. Top panel is for jets with an anti-$k_T$ radius $R=0.4$, while lower panel is for jets with $R=1$. }
\end{figure}

In the left panel of Fig.~\ref{fig:deltaeta} we show the $p_T$ density of the particles from the wake as a function of the azimuthal and rapidity separation with respect to the leading jet of a dijet system at the LHC with $\sqrt{s}=5.02$ ATeV, with $p_T^L>250$ GeV and $p_T^S>80$ GeV, with an angular separation in the transverse plane $\Delta \phi_D > 2\pi/3$ and an absolute difference in rapidity $|\Delta\eta_D|$. In the top left panel, the negative contribution from the more quenched subleading jet's QGP trough is not visible because it hits the wake of the leading jet, since $|\Delta\eta_D|$ is small, reducing the energy density in its surroundings. In the bottom left panel, when $|\Delta\eta_D|$ is large, the QGP trough of the subleading jet misses the leading jet, and therefore does not induce any energy loss.

We then construct a new observable, shown in the right panel of Fig.~\ref{fig:deltaeta}, that quantifies the amount of suppression of the leading jet, $R_{AA}^{\rm lead}$, in a dijet system as a function of the rapidity gap $|\Delta\eta_D|$ between the leading and the subleading jet. Without the contribution of the QGP trough, there is barely any dependence of $R_{AA}^{\rm lead}$ with $|\Delta\eta_D|$ \footnote{The visible drop at highest $|\Delta\eta_D|$ comes from the steepening of the spectrum once we require that both jets sit at relatively large rapidities.}. Once all particles are included, we see that there is notably more suppression at small $|\Delta\eta_D|$ than at large rapidity gaps, specially for the larger $R=1$ in the lower right panel. Given that most of the energy of a jet sits at its core, the curve including all contributions has a knee when $R\sim |\Delta\eta_D|$.

\section{Conclusions}

We have presented results for the angular dependence of jet suppression at RHIC and at LHC. We have discussed the role that the hydrodynamization of energy and momentum play in the description of such observable by sequentially analyzing the consequences of the different contributions that build up the total energy of the jet. Even though wider jets lose more energy, they can also recapture a larger fraction of the ``lost'' energy. These competing effects yield a radius dependence which is relatively flat, specially at the highest jet momenta at LHC (see~\cite{Pablos:2019ngg}), in resonance with what has been recently measured in experiments~\cite{Acharya:2019jyg,CMS:2019btm,Haake:2019pqd,Adam:2020wen}. We have observed that the diffusion wake of one jet can affect the other, inducing energy loss due to an over-subtraction effect, if the jets are relatively aligned in rapidity. This motivates the introduction of a new observable, namely the leading jet suppression as a function of the rapidity gap between the dijet system. This observable has the potential of unequivocally unravelling the fluid nature of quenched energy - a natural, and necessary consequence stemming from the hydrodynamic QGP paradigm.

\section*{Acknowledgements}

DP is supported by a grant from the Trond Mond Foundation with project number BFS2018REK01.


\begin{thebibliography}{99}

%\cite{Mehtar-Tani:2013pia}
\bibitem{Mehtar-Tani:2013pia}
Y.~Mehtar-Tani, J.~G.~Milhano and K.~Tywoniuk,
%``Jet physics in heavy-ion collisions,''
Int. J. Mod. Phys. A \textbf{28} (2013), 1340013
doi:10.1142/S0217751X13400137
[arXiv:1302.2579 [hep-ph]].
%125 citations counted in INSPIRE as of 04 Sep 2020

%\cite{Qin:2015srf}
\bibitem{Qin:2015srf}
G.~Y.~Qin and X.~N.~Wang,
%``Jet quenching in high-energy heavy-ion collisions,''
Int. J. Mod. Phys. E \textbf{24} (2015) no.11, 1530014
doi:10.1142/S0218301315300143
[arXiv:1511.00790 [hep-ph]].
%160 citations counted in INSPIRE as of 04 Sep 2020

%\cite{Acharya:2018zuq}
\bibitem{Acharya:2018zuq}
S.~Acharya \textit{et al.} [ALICE],
%``Anisotropic flow of identified particles in Pb-Pb collisions at $ {\sqrt{s}}_{\mathrm{NN}}=5.02 $ TeV,''
JHEP \textbf{09} (2018), 006
doi:10.1007/JHEP09(2018)006
[arXiv:1805.04390 [nucl-ex]].
%40 citations counted in INSPIRE as of 04 Sep 2020

%\cite{Casalderrey-Solana:2014bpa}
\bibitem{Casalderrey-Solana:2014bpa} 
  J.~Casalderrey-Solana, D.~C.~Gulhan, J.~G.~Milhano, D.~Pablos and K.~Rajagopal,
  %``A Hybrid Strong/Weak Coupling Approach to Jet Quenching,''
  JHEP {\bf 1410}, 019 (2014)
  Erratum: [JHEP {\bf 1509}, 175 (2015)]
 % doi:10.1007/JHEP09(2015)175, 10.1007/JHEP10(2014)019
  [arXiv:1405.3864 [hep-ph]].
  %%CITATION = doi:10.1007/JHEP09(2015)175, 10.1007/JHEP10(2014)019;%%
  %71 citations counted in INSPIRE as of 12 Jul 2018
  
  %\cite{Casalderrey-Solana:2015vaa}
\bibitem{Casalderrey-Solana:2015vaa} 
  J.~Casalderrey-Solana, D.~C.~Gulhan, J.~G.~Milhano, D.~Pablos and K.~Rajagopal,
  %``Predictions for Boson-Jet Observables and Fragmentation Function Ratios from a Hybrid Strong/Weak Coupling Model for Jet Quenching,''
  JHEP {\bf 1603}, 053 (2016)
 % doi:10.1007/JHEP03(2016)053
  [arXiv:1508.00815 [hep-ph]].
  %%CITATION = doi:10.1007/JHEP03(2016)053;%%
  %32 citations counted in INSPIRE as of 12 Jul 2018
    
  %\cite{Sjostrand:2014zea}
\bibitem{Sjostrand:2014zea} 
  T.~Sj\"ostrand {\it et al.},
  %``An Introduction to PYTHIA 8.2,''
  Comput.\ Phys.\ Commun.\  {\bf 191}, 159 (2015)
  doi:10.1016/j.cpc.2015.01.024
  [arXiv:1410.3012 [hep-ph]].
  %%CITATION = doi:10.1016/j.cpc.2015.01.024;%%
  %1710 citations counted in INSPIRE as of 23 Jul 2019
    
  %\cite{Chesler:2014jva}
\bibitem{Chesler:2014jva} 
  P.~M.~Chesler and K.~Rajagopal,
  %``Jet quenching in strongly coupled plasma,''
  Phys.\ Rev.\ D {\bf 90}, no. 2, 025033 (2014)
 % doi:10.1103/PhysRevD.90.025033
  [arXiv:1402.6756 [hep-th]].
  %%CITATION = doi:10.1103/PhysRevD.90.025033;%%
  %29 citations counted in INSPIRE as of 12 Jul 2018
  
  %\cite{Chesler:2015nqz}
\bibitem{Chesler:2015nqz} 
  P.~M.~Chesler and K.~Rajagopal,
  %``On the Evolution of Jet Energy and Opening Angle in Strongly Coupled Plasma,''
  JHEP {\bf 1605}, 098 (2016)
 % doi:10.1007/JHEP05(2016)098
  [arXiv:1511.07567 [hep-th]].
  %%CITATION = doi:10.1007/JHEP05(2016)098;%%
  %11 citations counted in INSPIRE as of 12 Jul 2018
  
   %\cite{Shen:2014vra}
\bibitem{Shen:2014vra} 
  C.~Shen, Z.~Qiu, H.~Song, J.~Bernhard, S.~Bass and U.~Heinz,
  %``The iEBE-VISHNU code package for relativistic heavy-ion collisions,''
  Comput.\ Phys.\ Commun.\  {\bf 199}, 61 (2016)
  doi:10.1016/j.cpc.2015.08.039
  [arXiv:1409.8164 [nucl-th]].
  %%CITATION = doi:10.1016/j.cpc.2015.08.039;%%
  %175 citations counted in INSPIRE as of 23 Jul 2019
  
  %\cite{Casalderrey-Solana:2018wrw}
\bibitem{Casalderrey-Solana:2018wrw} 
  J.~Casalderrey-Solana, Z.~Hulcher, G.~Milhano, D.~Pablos and K.~Rajagopal,
  %``Simultaneous description of hadron and jet suppression in heavy-ion collisions,''
  Phys.\ Rev.\ C {\bf 99}, no. 5, 051901 (2019)
  doi:10.1103/PhysRevC.99.051901
  [arXiv:1808.07386 [hep-ph]].
  %%CITATION = doi:10.1103/PhysRevC.99.051901;%%
  %2 citations counted in INSPIRE as of 23 Jul 2019
  
  %\cite{Chesler:2007an}
\bibitem{Chesler:2007an} 
  P.~M.~Chesler and L.~G.~Yaffe,
  %``The Wake of a quark moving through a strongly-coupled plasma,''
  Phys.\ Rev.\ Lett.\  {\bf 99}, 152001 (2007)
  doi:10.1103/PhysRevLett.99.152001
  [arXiv:0706.0368 [hep-th]].
  %%CITATION = doi:10.1103/PhysRevLett.99.152001;%%
  %105 citations counted in INSPIRE as of 23 Jul 2019
  
  %\cite{Cooper:1974mv}
\bibitem{Cooper:1974mv} 
  F.~Cooper and G.~Frye,
  %``Comment on the Single Particle Distribution in the Hydrodynamic and Statistical Thermodynamic Models of Multiparticle Production,''
  Phys.\ Rev.\ D {\bf 10}, 186 (1974).
  doi:10.1103/PhysRevD.10.186
  %%CITATION = doi:10.1103/PhysRevD.10.186;%%
  %839 citations counted in INSPIRE as of 23 Jul 2019
  
   %\cite{Casalderrey-Solana:2016jvj}
\bibitem{Casalderrey-Solana:2016jvj} 
  J.~Casalderrey-Solana, D.~Gulhan, G.~Milhano, D.~Pablos and K.~Rajagopal,
  %``Angular Structure of Jet Quenching Within a Hybrid Strong/Weak Coupling Model,''
  JHEP {\bf 1703}, 135 (2017)
 % doi:10.1007/JHEP03(2017)135
  [arXiv:1609.05842 [hep-ph]].
  %%CITATION = doi:10.1007/JHEP03(2017)135;%%
  %26 citations counted in INSPIRE as of 14 Feb 2018

  
  %\cite{Chen:2017zte}
\bibitem{Chen:2017zte} 
  W.~Chen, S.~Cao, T.~Luo, L.~G.~Pang and X.~N.~Wang,
  %``Effects of jet-induced medium excitation in $\gamma$-hadron correlation in A+A collisions,''
  Phys.\ Lett.\ B {\bf 777}, 86 (2018)
  doi:10.1016/j.physletb.2017.12.015
  [arXiv:1704.03648 [nucl-th]].
  %%CITATION = doi:10.1016/j.physletb.2017.12.015;%%
  %29 citations counted in INSPIRE as of 23 Jul 2019
  
  %\cite{Yan:2017rku}
\bibitem{Yan:2017rku}
L.~Yan, S.~Jeon and C.~Gale,
%``Jet-medium interaction and conformal relativistic fluid dynamics,''
Phys. Rev. C \textbf{97} (2018) no.3, 034914
doi:10.1103/PhysRevC.97.034914
[arXiv:1707.09519 [nucl-th]].
%5 citations counted in INSPIRE as of 04 Sep 2020

%\cite{Pablos:2019ngg}
\bibitem{Pablos:2019ngg}
D.~Pablos,
%``Jet Suppression From a Small to Intermediate to Large Radius,''
Phys. Rev. Lett. \textbf{124} (2020) no.5, 052301
doi:10.1103/PhysRevLett.124.052301
[arXiv:1907.12301 [hep-ph]].
%2 citations counted in INSPIRE as of 04 Sep 2020
  
  %\cite{Cacciari:2011ma}
\bibitem{Cacciari:2011ma} 
  M.~Cacciari, G.~P.~Salam and G.~Soyez,
  %``FastJet User Manual,''
  Eur.\ Phys.\ J.\ C {\bf 72}, 1896 (2012)
  doi:10.1140/epjc/s10052-012-1896-2
  [arXiv:1111.6097 [hep-ph]].
  %%CITATION = doi:10.1140/epjc/s10052-012-1896-2;%%
  %2942 citations counted in INSPIRE as of 28 Jul 2019
  
  %\cite{Milhano:2015mng}
\bibitem{Milhano:2015mng} 
  J.~G.~Milhano and K.~C.~Zapp,
  %``Origins of the di-jet asymmetry in heavy ion collisions,''
  Eur.\ Phys.\ J.\ C {\bf 76}, no. 5, 288 (2016)
  doi:10.1140/epjc/s10052-016-4130-9
  [arXiv:1512.08107 [hep-ph]].
  %%CITATION = doi:10.1140/epjc/s10052-016-4130-9;%%
  %52 citations counted in INSPIRE as of 05 Sep 2019
  
  %\cite{Rajagopal:2016uip}
\bibitem{Rajagopal:2016uip} 
  K.~Rajagopal, A.~V.~Sadofyev and W.~van der Schee,
  %``Evolution of the jet opening angle distribution in holographic plasma,''
  Phys.\ Rev.\ Lett.\  {\bf 116}, no. 21, 211603 (2016)
  doi:10.1103/PhysRevLett.116.211603
  [arXiv:1602.04187 [nucl-th]].
  %%CITATION = doi:10.1103/PhysRevLett.116.211603;%%
  %21 citations counted in INSPIRE as of 05 Sep 2019
  %\cite{Brewer:2017fqy}
  
  %\cite{Casalderrey-Solana:2019ubu}
\bibitem{Casalderrey-Solana:2019ubu}
J.~Casalderrey-Solana, G.~Milhano, D.~Pablos and K.~Rajagopal,
%``Modification of Jet Substructure in Heavy Ion Collisions as a Probe of the Resolution Length of Quark-Gluon Plasma,''
JHEP \textbf{01} (2020), 044
doi:10.1007/JHEP01(2020)044
[arXiv:1907.11248 [hep-ph]].
%13 citations counted in INSPIRE as of 04 Sep 2020
  
  %\cite{Caucal:2020xad}
\bibitem{Caucal:2020xad}
P.~Caucal, E.~Iancu, A.~H.~Mueller and G.~Soyez,
%``Nuclear modification factors for jet fragmentation,''
[arXiv:2005.05852 [hep-ph]].
%1 citations counted in INSPIRE as of 04 Sep 2020

%\cite{Adam:2020wen}
\bibitem{Adam:2020wen}
J.~Adam \textit{et al.} [STAR],
%``Measurement of inclusive charged-particle jet production in Au+Au collisions at $\sqrt{s_{NN}}$=200 GeV,''
[arXiv:2006.00582 [nucl-ex]].
%2 citations counted in INSPIRE as of 04 Sep 2020

%\cite{Acharya:2019jyg}
\bibitem{Acharya:2019jyg}
S.~Acharya \textit{et al.} [ALICE],
%``Measurements of inclusive jet spectra in pp and central Pb-Pb collisions at $\sqrt{s_{\rm{NN}}}$ = 5.02 TeV,''
Phys. Rev. C \textbf{101} (2020) no.3, 034911
doi:10.1103/PhysRevC.101.034911
[arXiv:1909.09718 [nucl-ex]].
%9 citations counted in INSPIRE as of 04 Sep 2020

%\cite{Haake:2019pqd}
\bibitem{Haake:2019pqd}
R.~Haake [ALICE],
%``Machine Learning based jet momentum reconstruction in Pb-Pb collisions measured with the ALICE detector,''
[arXiv:1909.01639 [nucl-ex]].
%4 citations counted in INSPIRE as of 04 Sep 2020

%\cite{CMS:2019btm}
\bibitem{CMS:2019btm}
 [CMS],
%``Measurement of Jet Nuclear Modification Factor in PbPb Collisions at $\sqrt{s_{NN}}$ = 5.02 TeV with CMS,''
CMS-PAS-HIN-18-014.
%5 citations counted in INSPIRE as of 04 Sep 2020

\end{thebibliography}
\end{document}